# Data Center Interconnects at 400G and Beyond

Michael H. Eiselt*, Annika Dochhan*, and Joerg-Peter Elbers**
* ADVA Optical Networking SE, Maerzenquelle 1-3, 98617 Meiningen, Germany
** ADVA Optical Networking SE, Fraunhoferstasse 9a, 82152 Martinsried/Munich, Germany

**Abstract**
**Current trends in Data Center Interconnectivity are considered in the light of increasing traffic and under the constraint of limited cost and power consumption.**

## I. INTRODUCTION

Worldwide data center IP traffic is growing by 25% per year, while the traffic between data centers even grows by 32.7% annually [1]. While the traffic is growing, cost and energy consumption of the network need to stay constant. Lowest cost solutions have already been implemented for data center interconnects over distances of 80-120 km by using intensity modulation and direct detection for channel data rates up to 100 Gb/s [2,3]. Further growth in this direction is, however, met by two stringent requirements: First, network installation and operation should be as simple as possible. This prohibits the use of chromatic dispersion compensation modules that need to be tailored to specific link distances. Second, to make optimum use of the fiber infrastructure, spectral efficiency, i.e. fiber capacity per unit bandwidth, needs to be maximized. While these two requirements might point to the use of coherent transmission systems also for shorter, inter data-center systems [4], further considerations of cost and power consumption swing the scale towards simpler interfaces using intensity modulation and direct detection (IM-DD).

In this paper, we will consider how far simple network operation and high capacity can be achieved with IM-DD interfaces and which gap will remain towards coherent interfaces.

## II. IM-DD MODULATION FORMATS

On-off keying (OOK) has long been the modulation format of choice for any short, medium and long-haul transmission. To enable a higher per-channel capacity while being limited by the electrical bandwidth of the devices, multi-level modulation has since been implemented in commercially available interfaces. Especially, 4-level pulse amplitude modulation (PAM-4) has been standardized in IEEE 802.3, in ITU-T G.695 and in ITU-T G.959.1 for reaches up to 10 km. To overcome chromatic dispersion penalties, operation of the standard interfaces has only been specified in the wavelength band around 1290 nm (O-band), where the chromatic dispersion is between -5 and +1 ps/nm·km. A per-channel data rate of 400 Gb/s is achieved using 8 parallel wavelengths, each operating at 50 Gb/s and separated by 800 GHz. A narrower wavelength spacing is prevented by the low chromatic dispersion, leading to non-linear interaction between the wavelengths. Increasing the wavelength count and introducing dense wavelength division multiplexing (DWDM) requires operation in the C-band (above 1530 nm), where the fiber chromatic dispersion coefficient is larger than 12 ps/nm·km in standard transmission fiber.

### A. Chromatic Dispersion Tolerance and Compensation

Increasing the channel data rate beyond 400 Gb/s will require per-wavelength rates of at least 100 Gb/s to avoid extreme cost and power consumption due to excessive number of optical interfaces. With PAM-4 modulation, however, a 56-GBd signal carrying a data rate of 112 Gb/s has a chromatic dispersion (CD) limit of approximately 50 ps/nm, if no means for CD tolerance is applied [5]. This corresponds to approximately 3 km of standard fiber in the C-band.

While fixed dispersion compensation modules appear to be not well suited to compensate for the fiber CD, as the correct deployment would require a very precise characterization of the CD and the administration of a large number of different dispersion compensation modules, the combination of fixed and tunable modules can automatically compensate the CD to within the required limit [6], thus coming close to the advantages of coherent systems.

Several DSP-based CD pre- and post-compensation techniques or IM/DD systems have been proposed recently, including the use of single side band (SSB) modulation for PAM-4 with non-linear mixing compensation [7] or Kramers-Kronig reception, but enhanced CD tolerance comes at the expense of increased hardware and DSP complexity [8,9].

### B. Spectral Efficiency

Depending on the operators of the DCI connection and their access to unlit transmission fiber, spectral efficiency of the transmission might be an issue, which could become more important than cost and power consumption. While coherent transmission systems are known to provide for a spectral efficiency of more than 5 b/s/Hz over DCI distances (e.g. 400 Gb/s transported using 16-QAM at 56 GBd within a 75-GHz spectral slot, yielding an SE of 5.33 b/s/Hz), the SE of IM-DD systems is typically on the order of 1 b/s/Hz. The use of SSB modulated PAM-4 can slightly increase this SE, e.g. to 1.33 b/s/Hz transporting 100 Gb/s at 56 GBd within a 75-GHz slot. Nevertheless, if a high spectral efficiency is required, as fiber cost is dominant, coherent transmission appears to be the format of choice.

## C. Noise Tolerance

A realistic fiber loss of 20 dB for an 80-km fiber link will require optical amplification before demultiplexing and detection of the DWDM spectrum. Therefore, the tolerance to optical noise, measured by the acceptable optical signal-to-noise ratio (OSNR) is the relevant parameter for DCI transmission, as opposed to the minimum received power tolerance for 10-km links. Assuming realistic system parameters with 6 dB amplifier noise figure, 3 dBm transmit power per channel and 3 dB margin, the achievable OSNR of a DCI link is limited to approximately 32 dB. This is marginally sufficient for 112 Gb/s per wavelength IM/DD transmission, where the lowest published results are in the range of 28 dB and higher [8-10]. The achievable link OSNR can be improved by employing more sophisticated, albeit operationally more complex, Raman amplification.

For coherent transmission, even 1 Tb/s per wavelength transmission has been experimentally demonstrated to require 31 dB of OSNR to achieve error free transmission with 100-GBd 64-QAM modulation and a high-performance FEC [11]. Current commercially available coherent transceivers require approximately 11 dB OSNR for 100 Gb/s per wavelength (DP-QPSK) and 19 dB OSNR for 200 Gb/s per wavelength (16-QAM) transmission. These lower OSNR values allow for a potentially lower-cost implementation of the optical link than required for IM/DD transmission.

## III. FURTHER INCREASE OF FIBER CAPACITY

An alternative and complement to increasing the spectral efficiency of the modulation format is the increase of the useable optical bandwidth. It has already been mentioned that the step to DWDM required the transition to the C-band to avoid non-linear inter-channel effects. Extending the spectrum to the L-band will double the fiber capacity. This approach was already proposed and implemented twenty years ago, when the capacity of long-haul systems was extended at the cost of additional band splitters and amplifiers, reducing the system reach performance and increasing the system cost. For DCI systems, however, amplification is less of an issue than for long-haul systems. The introduction of a 1.5-dB band splitter loss would reduce the system reach not by a factor of 1.4 (corresponding to an OSNR reduction of 1.5 dB), but rather by 6 km, corresponding to a fiber loss of 1.5 dB. For typically single-span systems, Stimulated Raman Scattering due to the wider optical bandwidth will only minimally increase. Multi-pump Raman gain could even be used for wide-band amplification to support the transition to the L-band and even beyond.

## IV. CONCLUSIONS

High data-rate transmission over data center interconnect distances can be based on IM-DD or coherent interfaces. While IM-DD has advantages in terms of equipment cost and power consumption, coherent transmission yields a higher spectral efficiency and, therefore, capacity per fiber. The final choice depends on the accessibility and cost of unlit fiber.

## ACKNOWLEDGMENT

This project has received funding from the European Union's Horizon 2020 research and innovation programme under grant agreement No 762055 (BlueSpace project) and from the German ministry of education and research (BMBF) under contract 16KIS0477K (SENDATE Secure-DCI project).